\documentclass[twocolumn,showpacs,preprintnumbers,amsmath,amssymb]{revtex4}
\usepackage{graphicx}% Include figure files
\usepackage{dcolumn}% Align table columns on decimal point
\usepackage{bm}% bold math

\begin{document}

%\preprint{rev. 12/16/04}

\title{Search for Correlated High Energy Cosmic Ray Events with CHICOS}% Force line breaks with \\

\author{
B.~E.~Carlson, E.~Brobeck, C.~J.~Jillings, M.~B.~Larson}
 \altaffiliation[Present address: ]{Center for Gravitational Wave Physics,
 Pennsylvania State University, University Park, PA 16802.}%Lines break automatically or can be forced with \\

\author{ T.~W.~Lynn}
\author{R.~D.~McKeown}%
\affiliation{%
W.~K.~Kellogg Radiation Laboratory, California Institute of Technology, Pasadena, CA 91125, USA
%\\
}%

\author{James E.~Hill}
\affiliation{
Department of Physics, California State University, Dominguez Hills, Carson, CA 90747, USA
}%
\author{B.~J.~Falkowski and R.~Seki}
\affiliation{ Department of Physics and Astronomy, California State University, Northridge, CA 91330, USA
}%

\author{J.~Sepikas}
\affiliation{
Department of Astronomy, Pasadena City College, Pasadena, CA 91106, USA
}%
\author{G.~B.~Yodh}
\affiliation{ Department of Physics and Astronomy, University of California, Irvine, CA 92697-4575, USA
}%
\date{\today}% It is always \today, today,

\begin{abstract}
We present the results of a search for time correlations in high energy cosmic ray data (primary $E >
10^{14}$~eV) collected by the California HIgh school Cosmic ray ObServatory (CHICOS) array. Data from 60
detector sites spread over an area of 400~km$^2$ were studied for evidence of isolated events separated by more
than 1~km with coincidence times ranging from 1~$\mu$sec up to 1~second.  The results are consistent with the
absence of excess coincidences except for a $2.9 \sigma$ excess observed for coincidence times less than
10~$\mu$sec. We report upper limits for the coincidence probability as a function of coincidence time.

\end{abstract}
%Valid PACS numbers may be entered using the \verb+\pacs{#1}+ command.
\maketitle

%\section{\label{sec:level1}First-level heading:\protect\\ The line
%break was forced \lowercase{via} \textbackslash\textbackslash}

The spectrum of cosmic rays above $10^{15}$~eV has been studied up to energies of $10^{20}$~eV by observing the
large extensive air showers created by the primary incident particles \cite{nagano}. Below $10^{15}$~eV there
have been direct measurements via instruments flown in satellites or as balloon payloads \cite{direct}. The
energy spectrum falls steeply with a power law $E^{-2.7}$ for energies up to the ``knee'' at $4 \times
10^{15}$~eV, and even more steeply, $E^{-3.0}$, beyond the knee. Shock acceleration in supernovae provides a
successful explanation up to the knee, while the source of particles beyond the knee region is speculative.
Above $10^{19.6}$~eV, extra-galactic protons should interact with the cosmic microwave background (CMB) and
thereby lose energy, resulting in a sharp decrease in the number of cosmic rays with energies above the
Greisen-Zatsepin-Kuzmin (GZK) limit of $10^{19.6}$~eV \cite{greisen}. Experimental data regarding the existence
of the GZK cutoff are inconclusive at this point \cite{AGASA,HiRes}.

Correlations between cosmic rays would indicate that the particles have some common history.  Such correlations
could provide information about the source of the cosmic rays, the number or distribution of sources, or about
the propagation of cosmic rays. For example, a recent study of ultra-high energy cosmic rays by AGASA
\cite{agasaclustering} indicates clustering of the directions of origin in the sky.

This paper addresses the possibility that isolated cosmic ray events separated by $> 1$~km arrive in time
coincidence. Such correlated cosmic ray events could result, for example, from the photodisintegration of heavy
nuclei (\emph{i.e.}, iron) by solar photons \cite{photodisintegration}. A previous study,  \cite{carrelmartin},
indicated episodic evidence for time correlations up to $10^{-3}$~s in events separated by $\sim 100$~km. A more
recent search for correlated events \cite{LAAS} at large distance scales, $\sim 500$~km, found a few candidate
events but these were consistent with interpretation as accidental coincidences between uncorrelated events. We
have studied 17 months of data obtained with the CHICOS array during January 2003 through July 2004 and searched
for evidence of correlated air shower events separated by $> 1$~km with energy threshold $10^{14}$ eV. This
paper reports the results of that search.

\section*{CHICOS}

\begin{figure}
\begin{center}
%\vspace{2.5 in}
\includegraphics[clip=true,scale=.4]{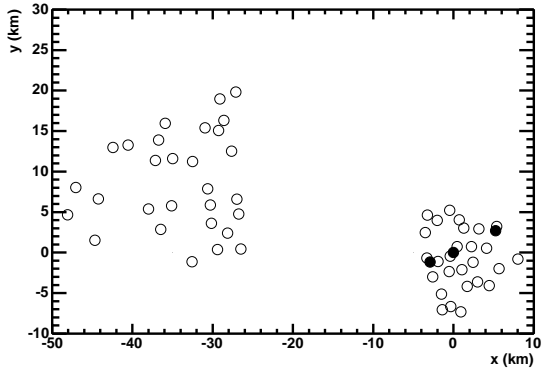}
%% \plotone{plots/expodist_nofit_pub.epsi}
\end{center}
\caption{Locations of the 60 operating CHICOS sites used in the analysis in this paper are shown as open
circles. The Caltech site is located at the origin, $x$ indicates distance in the easterly direction, and $y$
indicates the distance in the northerly direction. The sites in the San Gabriel valley are clustered around the
origin, whereas the sites in the San Fernando valley are centered about 35~km to the west. The filled circles
are not included in the $>1$~km analysis.
%The two sites
%indicated by stars (rather then squares) are removed from the some analyses to insure site spacing greater than
%1~km (see text).
} \label{fig:map}
\end{figure}

The California HIgh school Cosmic ray ObServatory (CHICOS) observes cosmic ray induced air showers with an array
of detector sites located on school roofs in the Los Angeles area (Lat.~$34.1^\circ$, Long.~$-118.1^\circ$,
average 245~m above sea level). The sites are separated by distances of typically $2-3$ kilometers, with the
overall array covering an area of $\sim 400$~km$^{2}$. As shown in Fig.~\ref{fig:map}, the detector sites are
located in two major groups in the San Gabriel and San Fernando valleys, separated by $\sim 35$~km. These 60
sites contain four pairs separated by less than 1~km, and all other pairs of sites have larger separations.
During the period corresponding to the dataset reported in this paper, the number of operational sites increased
from 31 to 60.

Every detector site contains two plastic scintillator detectors, separated by $\sim 3$ meters, with each
detector having $\sim 1$~m$^2$ area and $5-10$~cm thickness. Photomultipler signals are processed by a
custom-built time-over-threshold discriminator circuit. Timing of the detector signals and the GPS receiver
signals is facilitated by use of National Instruments 6602 80MHz timer/counter cards.  We have verified that the
GPS receiver, either Motorola UT+ or M12, provides relative timing accuracy of $\sim 50$~nsec. Pulse heights,
determined from the time-over-threshold measurements, are calibrated and continuously monitored using the high
flux ($\sim 200$~Hz per detector) of incident throughgoing muons. The discriminator threshold is set at $\sim
1/4$ of the observed muon peak.

Data are stored on local hard disk and automatically transferred to Caltech via internet every night by the
computer located at each site. ``Trigger'' events are defined as those where both detectors at a site record
signals greater than 2 single vertical particles. Each site generates about 1000 trigger events per day. These
``trigger'' events and all ``match'' hits (single-detector hits within $50$~$\mu$sec of any trigger elsewhere in
the array) are transferred to Caltech. Shower-building software on our server then analyzes these data every
morning and constructs files of built showers for further analysis.

Extensive air showers generate coincidences among several sites. However, the core of the air shower that will
generate trigger events is generally much smaller than the spacing between CHICOS sites. Therefore, a typical
air shower event involves one trigger with ``match'' events at the neighboring sites. The trigger events are
used to locate the cores of potential air showers, where the particle density is relatively high. Showers that
trigger one site with matching hits at several neighboring sites must have extremely high energies of $>
10^{18}$~eV.  We do observe such large air shower events with the CHICOS array, presently at a rate of about one
per month. The arrival times and intensities of detector signals provide the information necessary to
reconstruct the incident direction and energy of the primary particle that created the air shower.

Most trigger events are isolated single events ({\it i.e.}, no nearby matches) which are generated by much more
frequent lower energy showers with a threshold of about $10^{14}$~eV. The rate of these triggers is comparable
to expectations based on the previously measured flux \cite{nagano} and computer simulations of air showers with
the AIRES code \cite{AIRES}. In this paper we study double trigger events where two sites separated by more than
1~km both record trigger events within a certain coincidence time.
%(Single air shower events do not
%typically cause such events, but there is the possibility that rare fluctutations could cause them. However, we
%note that if separate correlated showers do occur then they would certainly generate the type of signal that we
%are seeking.)

In a separate study, we deployed 4 additional sites (not shown in Fig.~1) separated by $100-400$~m on the
Caltech campus to provide us with a more abundant sample ($\sim 10$ per day) of lower energy showers in the
range $10^{15}-10^{18}$~eV. These data give us confidence that the CHICOS equipment functions as expected,
exhibiting the correlated data that should result from air showers. In particular, the shower properties can be
reconstructed and the events are well described by single air showers. These data are under analysis and will be
reported in a future publication. We note that in this dataset, double triggers are not uncommon features of
larger air showers with a core size capable of generating triggers at two sites separated by $< 600$~m. The rate
of these double triggers falls steeply with distance between sites, and is well described by AIRES simulations.
In this paper we are studying the possibility of double triggers at sites separated by $>1$~km, so data from
these additional 4 sites on the Caltech campus are {\it not} used in the analysis in this paper.

\begin{figure}
\includegraphics[clip=true,scale=.32,angle=-90]{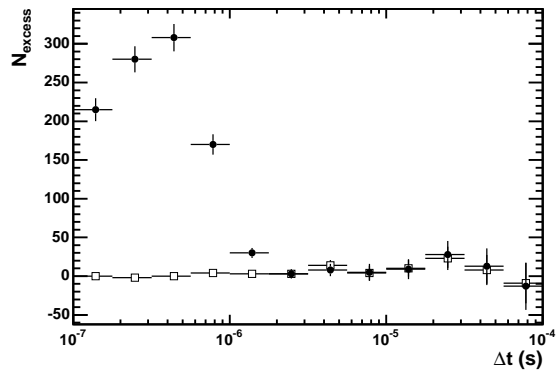}
%% \plotone{plots/expodist_nofit_pub.epsi}
\caption{Histogram of the excess of successive trigger pairs vs. the time difference. The data show a large
signal associated with the tails of air showers (solid circles) that disappears when the distance between the
pairs is required to be greater than 1~km (open squares).} \label{fig:fig1}
\end{figure}

\section*{Correlation Analysis}

The main focus of this analysis has been the trigger data sequence, which forms a complete record of all the
events detected by the array with sufficient energy to trigger a single site.   In order to examine these data
for time correlations, a randomized data set was constructed  directly from the real data, as in
\cite{carrelmartin}, by offsetting the sequence of triggers at each site by some integer number of seconds
relative to the other sites.  Since a shift of several seconds is small compared to the time for drift in the
average trigger rate, the randomized data should reproduce all aspects of the real data associated with
accidental coincidences. Deviations of correlations observed in the real data relative to the randomized data
could be indications of real correlations (\emph{i.e.}, not accidental) in the data.

\begin{figure}[b!]
\includegraphics[clip=true,scale=.32,angle=-90]{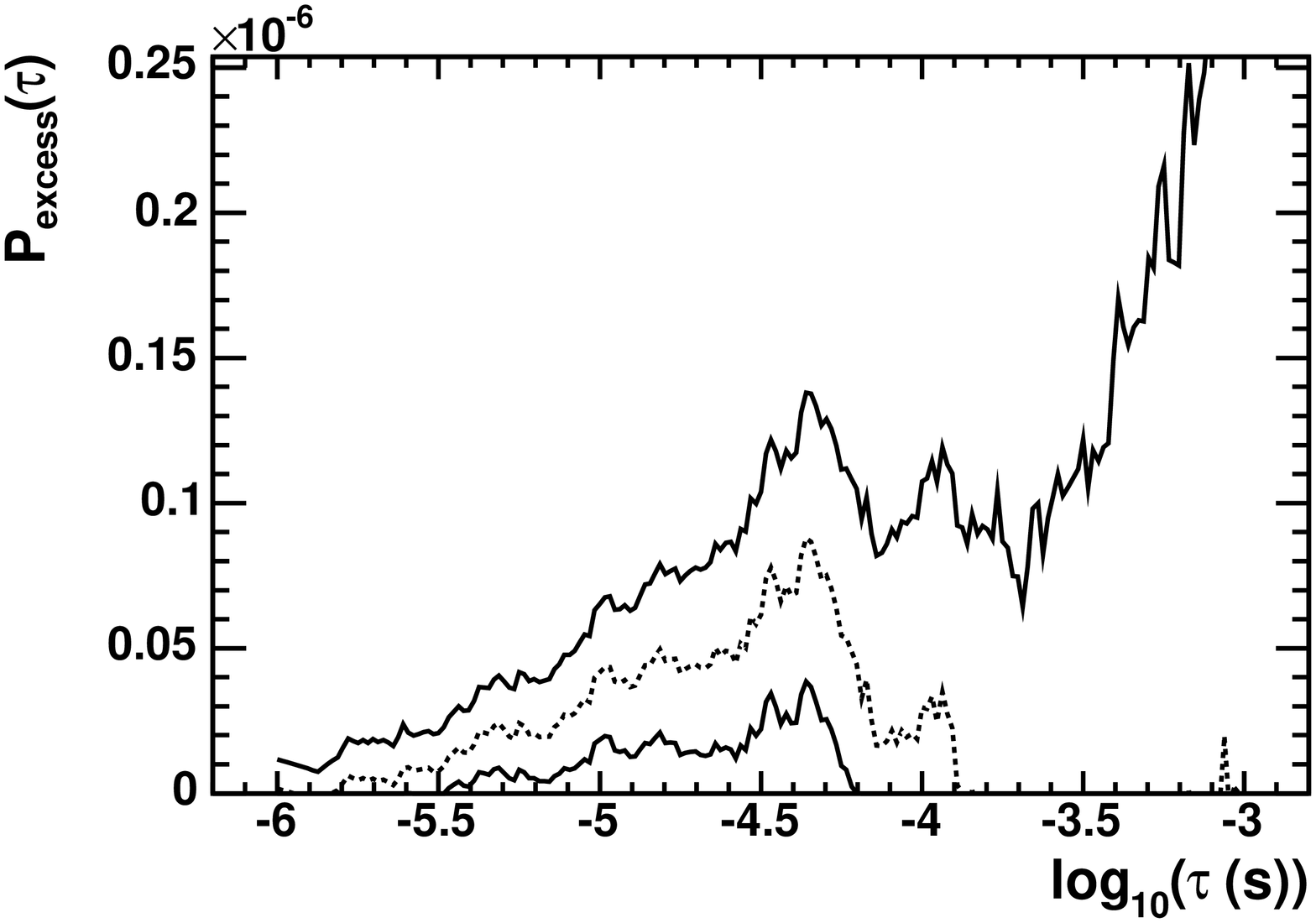}
\includegraphics[clip=true,scale=.32,angle=-90]{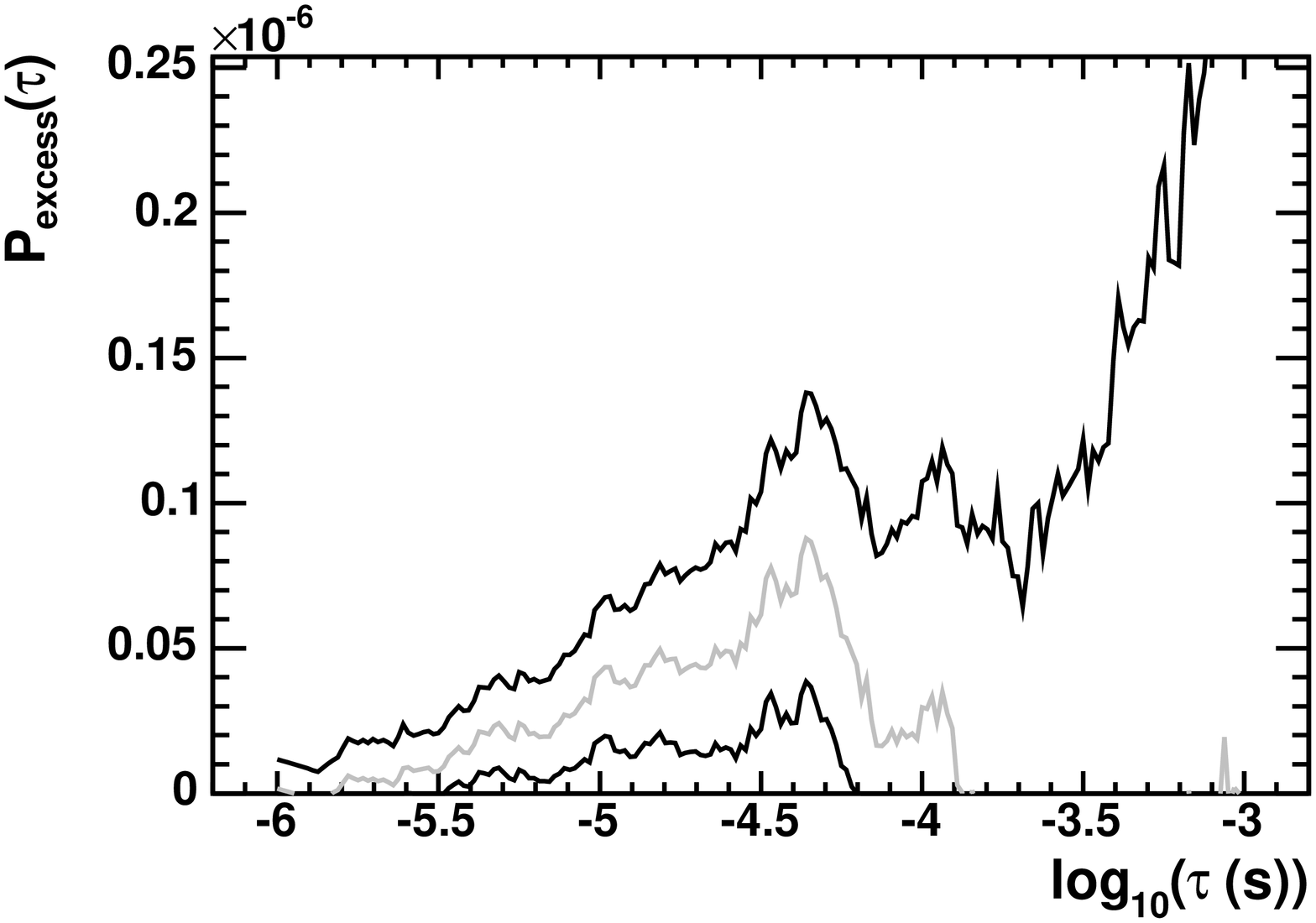}
%% \plottwo{plots/FCCI_prob_short.pub.epsi}{plots/FCCI_prob_long.pub.epsi}
\caption{Probability of an excess trigger coincidence per site plotted vs. cumulative coincidence time scale.
The upper plot shows the results for short time scales $\tau< 10^{-3}$~s and the lower plot shows the longer
time scales $10^{-3}<\tau< 1$~s. The lines drawn are the upper end (solid), center (dotted) and lower end
(solid) of a 90\% Feldman-Cousins confidence interval. The upper curve is interpreted as an upper limit for the
probability per site of real coincidences for the corresponding time scale.} \label{FCCIs}
\end{figure}

During the 17 month period, $1.68 \times 10^{7}$ trigger events were selected (according to criteria described
below) and analyzed for time correlations. The distribution of time differences between consecutive triggers
falls exponentially as expected, but does not follow an exact exponential distribution due to the varying number
of operating detector sites. For a coincidence time interval $\Delta t$ we count the number of consecutive
trigger pairs with time separation less than $\Delta t$ and subtract the corresponding number of pairs in the
randomized data to form an excess $N_{\rm excess}$, which may be positive or negative. If both members of a
successive pair are from the same site, that pair is not counted (to eliminate instrumental effects such as PMT
afterpulsing). Fig.~\ref{fig:fig1} shows $N_{\rm excess}$ as a function of the coincidence time $\Delta t$. The
large signal corresponding to real air showers is evident at short time intervals $\Delta t < 2 ~\mu$s. If 3 of
the 60 sites are omitted (see Fig.~1), all pairs with separation 1~km or less are removed from the data set and
the air shower signal disappears. In the following, we select sites so that relative distances are always
greater than 1~km.

In order to search for correlations on any time scale less than 1 second, we
%developed a method for scanning the
%time range to give the optimum sensitivity at each hypothesized coincidence time.  The
compute the
probability of an excess coincidence per site
%is computed
for the cumulative time interval $\{ 0 ,
\tau \}$ according to
\begin{equation}
P_{\rm excess} (\tau) = \frac {N_{\rm excess} (\Delta t< \tau ) } {N_{\rm trig}\, (\langle N_{\rm sites} \rangle
- 1)}
\end{equation}
in which $\langle N_{\rm sites} \rangle =37 $ is the average number of operational sites and $N_{\rm trig}$ is
the total number of trigger pairs. We compute a 90\% confidence interval for $P_{\rm excess}(\tau)$ using the
method in \cite{feldmancousins}, and interpret the upper limit as the 90\% confidence level upper limit for the
excess probability per site for the interval $\{ 0 , \tau \}$. The results are shown in Fig.~\ref{FCCIs}.

\begin{figure}
\includegraphics[clip=true,scale=.32,angle=-90]{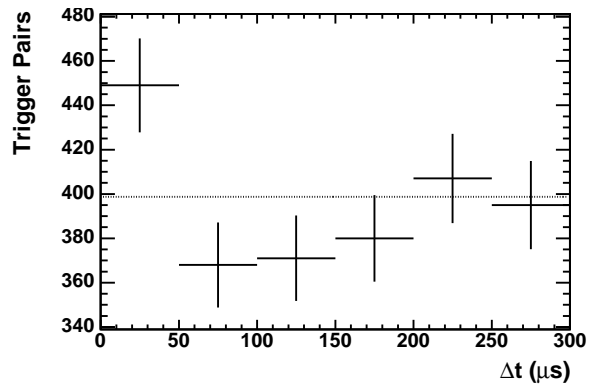}
%% \plotone{plots/timediffs_realandrandom_50usbins_randexpectations.pub.epsi}
\caption{Trigger pair time distribution for $\Delta t < 300 ~\mu$s.  Plotted points with error bars are real
data and the horizontal dotted line indicates the expectation from random coincidences.} \label{timediffs}
\end{figure}

The upper curve in Fig.~\ref{FCCIs} should grow as the $\sqrt{\tau}$, and it does increase in a fashion that is
consistent with expectations. One expects that the lower 90\% C.L. should behave in a symmetric manner, and
remain negative or close to zero. However, Fig.~\ref{FCCIs} indicates a rather substantial excess at time scales
of order $\sim 10^{-4.5}$~s, or $\sim 30 ~\mu$s. The excess heals itself at larger $\tau$ as we add much more
data that shows (apparently) no correlations. A less significant excess is perhaps evident at $\sim 0.1$~s.

\begin{figure}
\includegraphics[clip=true,scale=.32,angle=-90]{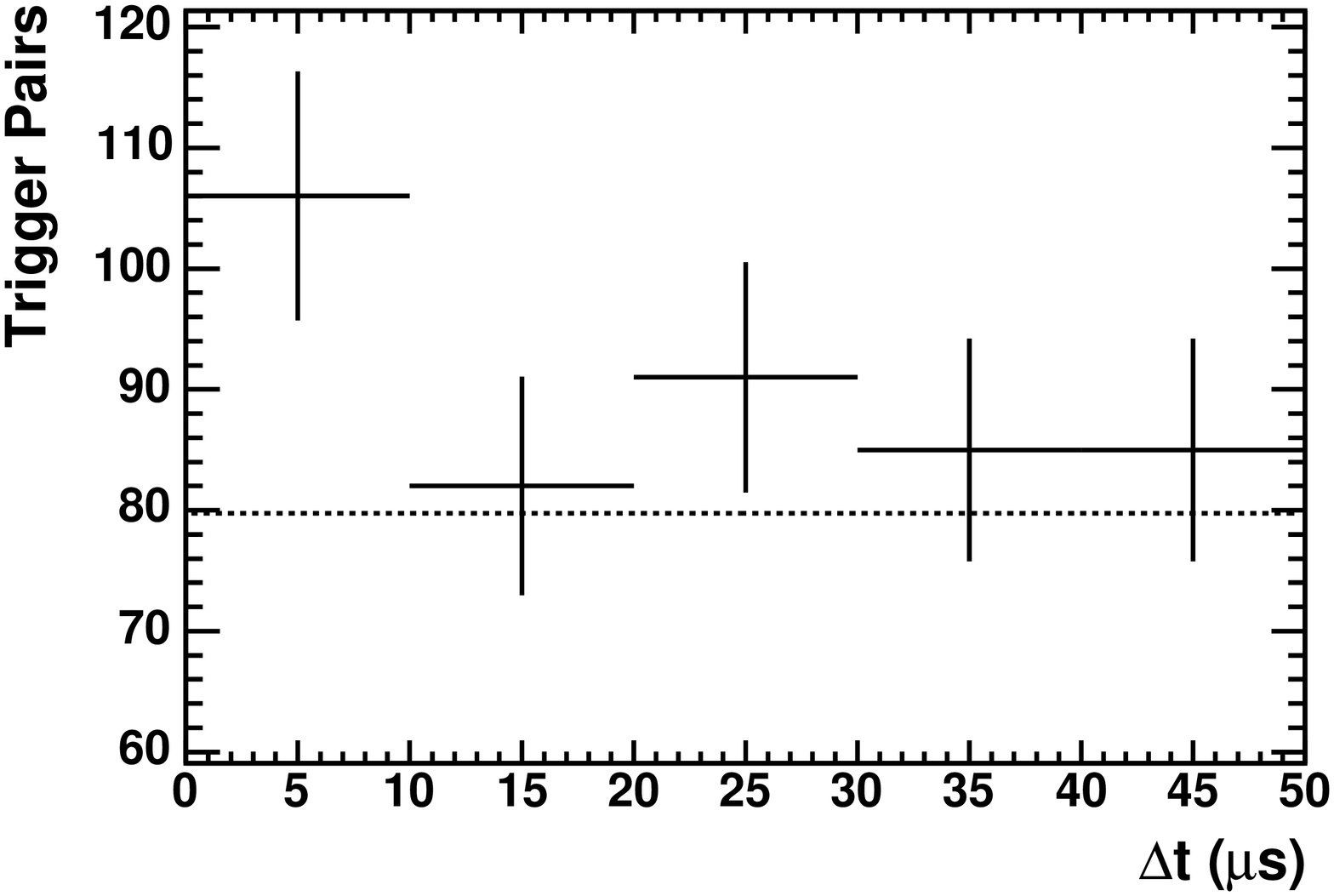}\\
\vspace*{0.2in}
\includegraphics[clip=true,scale=.32,angle=-90]{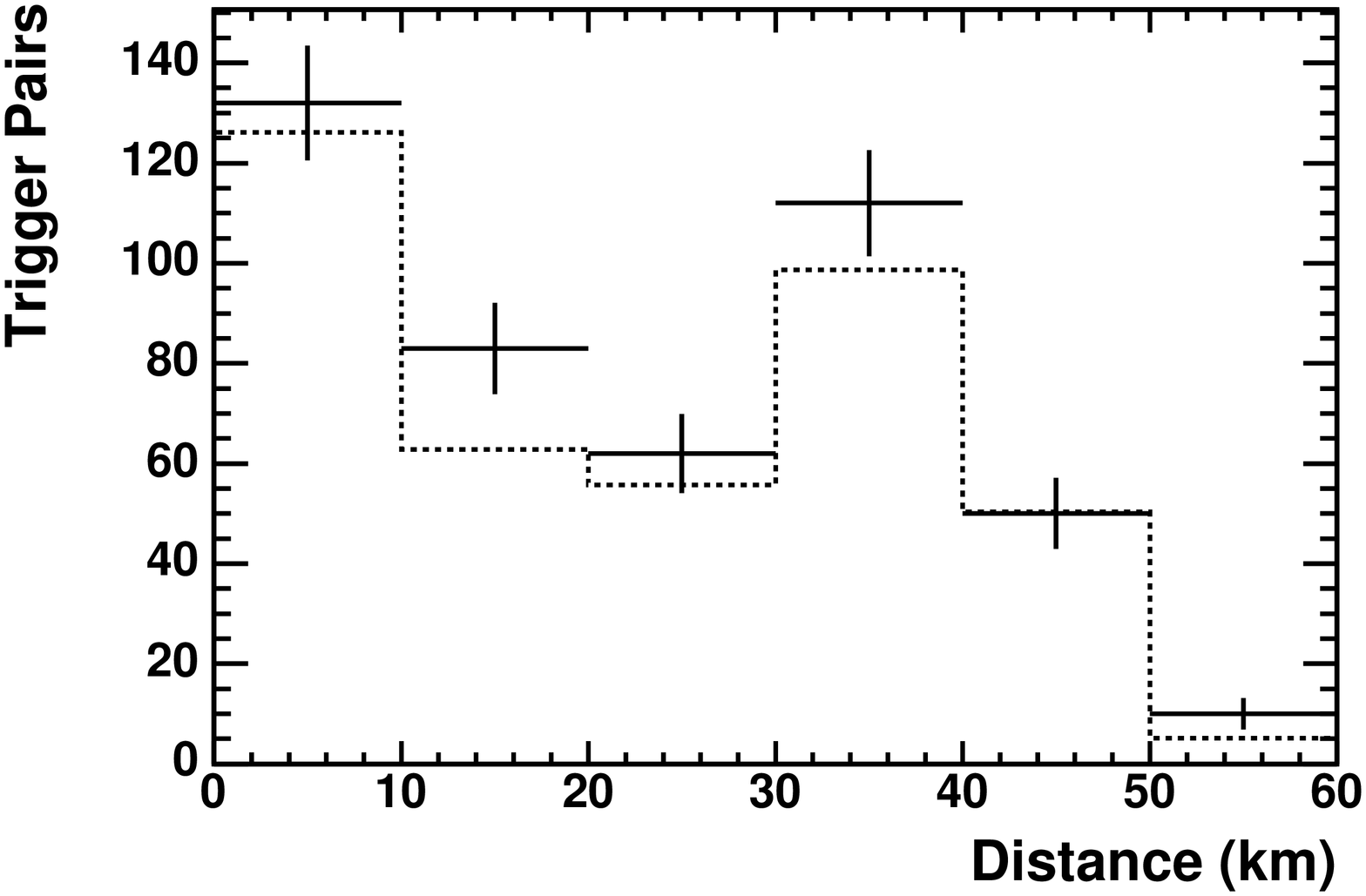}
%% \plottwo{plots/timediffs_realandrandom_10usbins_randexpectations.pub.epsi}{plots/dists_realrandom.3.epsi}
\caption{Time and distance distributions for coincidences separated by less than 50~$\mu$s. Real data is shown
as points with error bars, randomized data by dotted lines. The distance distributions for both the real and
randomized data are related to the spatial distribution of detector sites in Fig.~\ref{fig:map}, but the
difference between the real and random histograms potentially contains information on the excess coincidences. }
\label{timeandspace}
\end{figure}

A histogram of time differences for real data is plotted along with random expectations for short time scales in
Fig.~\ref{timediffs}, which indeed shows an excess in the $0-50~\mu$s bin as one might expect from
Fig.~\ref{FCCIs}. The random data are averaged over 1~ms and plotted as a constant line to improve the
statistics. The excess in the first bin is $2.5 \sigma$ over the expectation, significant at 99.26\% C.L.
%a positive result with 99.26\% C.L.

In Fig.~\ref{timeandspace} we display the data for the first bin of Fig.~\ref{timediffs}, \emph{i.e.}, $\Delta t
< 50 ~\mu$s. Although there is a preference for shorter time differences $\leq 10~\mu$s, no clustering at short
distances is evident in the spatial distribution, indicating the excess events are likely not due to single air
shower events. The excess in the first bin at $\Delta t < 10~ \mu$s is significant at 99.70\% C.L.

%\section*{Match Analysis}

The excess coincident trigger events (relative to random coincidence expectation) within the $0-50~\mu$s time
bin appear to be distributed over distances larger than $10$~km. The average separation distance for events in
the $0-50~\mu$s time bin is $L_{\rm data}= 23.1 \pm 0.7$~km, compared to $L_{\rm random} = 22.6 \pm 0.16$~km for
the randomized distribution. Thus the apparent excess events exhibit an average separation distance of $L_{\rm
excess} = 27 \pm 7$~km, consistent with $L_{\rm random}$. In contrast, we note that single air showers generated
by primary cosmic rays near vertical incidence are much smaller, extending only over several km, and should
exhibit a very steeply falling distribution in separation distance.

Since the apparent excess in our data is associated with average coincidence time $\Delta t \sim 17 ~\mu s$ and
$L \sim 27$~km these events would be consistent with rather vertical incidence $\theta \sim 15^{\circ}$ for the
primaries (or pairs of primaries). Highly inclined or horizontally incident single-shower events would imply
much longer average $\Delta t = L/c \simeq 90 ~\mu$s than indicated by the data in Fig.~4.

Finally, we note that we also attempted to find correlations of single isolated ``match'' events with distant
trigger events. Due to the high rate of accidental ``match'' hits in that analysis, the results were much less
sensitive and so are not reported here.

\section*{Summary and Conclusion}

A search for time correlations in cosmic ray data collected by the CHICOS project has been performed. The
results are generally consistent with a lack of any real correlation between isolated events, except for trigger
events ($E>10^{14}$~eV) with coincidence times $\Delta t < 50~\mu$s where an excess number of coincidences with
a significance of $2.5\sigma$ is observed. This excess is distributed uniformly over the 17 month data period,
and the events are randomly distributed over the array sites within the 400~km$^2$ area of the CHICOS array. For
smaller times $\Delta t < 10~\mu$s the data indicate an even more significant excess of $2.9\sigma$. We know of
no previous cosmic ray experiment that would be sensitive to correlated shower events at these energies
separated by $10-50$~km. The LAAS experiment \cite{LAAS} has only 8 sites at much larger distances ($\sim
500$~km) and would have seen $<1$ event given the rate we observe (assuming the correlations persist to those
larger distance scales).

If the apparent excess of correlated trigger events at $\Delta t < 50 ~\mu$s is interpreted as a real signal,
then we observe a coincidence probability of $P_{\rm excess} \sim 8\times 10^{-8}$ per site for each trigger
signal. This low probability implies that observing triple coincidences is extremely unlikely. However,
normalizing for the aperture of the detector arrangement at each site, one would then infer that a substantial
fraction, perhaps $10^{-3}$ or more, of high energy cosmic rays come in pairs or multiplets within a $\sim
1000$~km$^2$ area. However, this interpretation is strongly energy dependent, and since we have not measured the
event energies this could be true over any subset of energies in the range $10^{14}- 10^{18}$~eV.

Correlated events at $\sim 27$~km separations would probably result from rather local events, on the distance
scale of the solar system. Sources at galactic distance scales (or greater) might generate correlated events
over much larger distance scales, but it is extremely unlikely that particles could be reliably propagated at
such short transverse distances over $\sim 10^{17}$~km.
%Outside the earth's atmosphere in the solar system, the
%candidates for interaction of cosmic rays are photons from the sun and the solar wind. The solar wind is very
%dilute relative to the atmosphere.
The photodisintegration of heavy nuclei (\emph{i.e.}, iron) by solar photons is a possibility that has been
previously considered \cite{photodisintegration}, but the predicted fluxes are extremely low. Based on the
estimates in \cite{photodisintegration}, we expect that about $10^{-5}$ of the incident cosmic rays at $E \sim 6
\times 10^{17}$~eV will arrive as pairs within the size of the CHICOS array due to photodisintegration. These
pairs would typically consist of a heavy nucleus along with a $\sim 10^{16}$~eV nucleon. The rate of these pairs
within the 400 km$^2$ area of CHICOS would be about 1 per year. The efficiency for the $\sim 10^{16}$~eV nucleon
to generate a trigger at any of the 60 CHICOS sites is quite small, $\sim 6 \times 10^{-3}$, so the apparent
excess coincidence rate ($\sim 40$ events per year) is about $10^4$ times greater than the expected rate for
this process. Thus if we maintain that the correlations originate in the solar system, we probably need either
anomalous interactions with the solar wind or with solar photons.

The distance distribution of excess events in Fig.~\ref{timeandspace} yields an average separation distance
$L_{\rm excess} = 27 \pm 7$~km, much larger than extensive air showers from single primary particles near
vertical incidence. Although it seems unlikely, it may be that the excess coincidences are associated with
subtle unknown properties of large air shower events. It may be possible to explore such a possibility by
generating unthinned simulations with high spatial resolution ($\sim 1$~m) of air showers over a large $\sim
1000$~km$^2$ area. Although this is a formidable computational task for present simulation programs such as
AIRES \cite{AIRES}, we will attempt to address this issue in the near future. Another unlikely possibility is
that the generation of correlated shower events in the upper atmosphere separated by $>10$~km at ground level
could signal the onset of exotic new phenomena.
%The direct measurements at $E<10^{15}$~eV \cite{direct} would probably have revealed
%production of correlated showers if this were the appropriate energy range, so one would need to consider higher
%energies. The generation of $E> 10^{15}$~eV showers separated by $\sim 30$~km may require anomalous processes
%high in the atmosphere with very large transverse momenta ($> 100$~TeV). Such large transverse momenta could
%imply anomalous interactions or new heavy particles. (Since for incident protons or nuclei there is insufficient
%CM energy, the incident particles themselves would need to be very heavy, $\sim 1000$~TeV).

It is clearly desirable to obtain more data to improve the statistical precision and attempt to verify or refute
this observation. We hope to continue to expand and operate the CHICOS array to obtain a larger data set over
the next few years. In addition, it may be possible for the Auger observatory to search for such correlated
events. Further simulation studies of the properties of extensive air showers at large distances would also be
helpful in the interpretation of these data.

\section*{Acknowledgements}

%\begin{acknowledgments}
We are grateful for the generous support of Caltech and the Weingart Foundation in initiating the CHICOS
project. The donation of the detectors by the CYGNUS collaboration is gratefully acknowledged. Support from the
NSF (grants PHY-0244899 and PHY-0102502) and the donation of computers for the project by IBM Corporation are
also acknowledged. The volunteer efforts of many high school and middle school teachers
\footnote{http://www.chicos.caltech.edu/collaboration/collaboration-list.html} have been essential in the
deployment and operation of the CHICOS array, and we are delighted to acknowledge their participation.
Assistance from M.~Takeda and S.~Ho in generating AIRES simulations is gratefully acknowledged.
%\end{acknowledgments}

\end{document}